\documentclass[manuscript,screen]{acmart}
\AtBeginDocument{%
  }

\setcopyright{acmlicensed}
\copyrightyear{2018}
\acmYear{2018}
\acmDOI{XXXXXXX.XXXXXXX}
\acmConference[Conference acronym 'XX]{Make sure to enter the correct
  conference title from your rights confirmation email}{June 03--05,
  2018}{Woodstock, NY}
\acmISBN{978-1-4503-XXXX-X/2018/06}

\usepackage{multirow}

\begin{document}

\title{The Influence of Prior Discourse on Conversational Agent-Driven
Decision-Making}


\newcommand{\vnote}[1]{{\todo[inline, color=yellow!65]{Vivek: #1}}}
\newcommand{\snote}[1]{{\todo[inline, color=orange!65]{Stephen: #1}}}

\newcommand{\vtext}[1]{{\color{red}{#1}}}
\newcommand{\stext}[1]{{\color{blue}{#1}}}
\newcommand{\polish}[1]{{\color{orange}{#1}}}



\author{Stephen Pilli}
\email{stephen.pilli@ucdconnect.ie}
\orcid{0000-0003-1655-1782}
\affiliation{%
  \institution{University College Dublin}
  \city{Dublin}
  \country{Ireland}
}

\author{Vivek Nallur}
\orcid{0000-0003-0447-4150}
\email{vivek.nallur@ucd.ie}
\affiliation{%
  \institution{University College Dublin}
  \city{Dublin}
  \country{Ireland}
}

\renewcommand{\shortauthors}{Pilli et al.}

\begin{abstract}

Persuasion through conversation has been the focus of much research. Nudging is a popular strategy to influence decision-making in physical and digital settings. However, conversational agents employing ``nudging'' have not received significant attention. We explore the manifestation of cognitive biases—the underlying psychological mechanisms of nudging—and investigate how the complexity of prior dialogue tasks impacts decision-making facilitated by conversational agents. Our research used a between-group experimental design, involving 756 participants randomly assigned to either a simple or complex task before encountering a decision-making scenario. Three scenarios were adapted from Samuelson\'s classic experiments on status-quo bias, the underlying mechanism of default nudges. Our results aligned with previous studies in two out of three simple-task scenarios. Increasing task complexity consistently shifted effect-sizes toward our hypothesis, though bias was significant in only one case. These findings inform conversational nudging strategies and highlight inherent biases relevant to behavioural economics.

\end{abstract}


\keywords{Do, Not, Us, This, Code, Put, the, Correct, Terms, for,
  Your, Paper}

\received{20 February 2007}
\received[revised]{12 March 2009}
\received[accepted]{5 June 2009}

\maketitle

\section{Introduction}
\label{sec:Introduction}


To say that computers permeate every aspect of our modern life, is now trite. Digital interfaces mediate interactions with loved ones, transport, medical checkups, entertainment, and sometimes even food and sexual choices. The function and form of these interfaces influence not only our view of the world, but also \textit{how} we influence the world. Our decisions on which action to take, which option to ignore, and what aspect of a problem to pay attention to, are all affected by the contextual elements within which the decision scenario appears. When these contextual elements are created, enhanced, elided, magnified, or deleted by digital interfaces, it would be reasonable to say (we hope) that the digital interface is critically affecting the decision itself.  \\

In its simplest form, decision-making requires a decision scenario, and set of alternatives to choose from. The presentation of alternatives invariably influences the decision-maker, and therefore even a random or `unthoughtful' presentation creates an impact (whether intended or not). This has led to terms like `choice architecture' being used to characterise the presentation of alternatives in decision scenarios~\cite{thaler_choice_2010}. An etiology of decision-making reveals that the alternatives we pay attention to, are impacted by the cognitive biases that all human beings have~\cite{VanEyghen_2022}. These have often been called shortcuts in decision-making~\cite{kahneman_thinking_2011}. At this juncture, it is important to note that these shortcuts are not necessarily good or bad, in a traditional sense of evaluating decisions. Rather they pertain to the speed with which certain pathways in the brain get activated in particular decision-making contexts. These shortcuts arise from human experiences, emotions, beliefs, and preferences, with different kinds of decisions being affected by different biases. Other common factors include cognitive limitations (such as memory), information limitation, emotions, feelings, and social norms~\cite{eidelman_bias_2012}. Over a hundred cognitive biases have been catalogued over the years, in the field of behavioural economics~\cite{carter2007behavioral}. For example, \textit{status quo bias} is the preference for the maintenance of one\'s current or previous state of affairs, or a preference to not undertake any action to change this current or previous state~\cite{samuelson_status_1988}. An instance of where such a bias has been used in the real world is the mobile payment app Square, where customers are required to actively choose a ``no tipping'' option if they decide not to leave a tip. Such a choice architecture has resulted in increased tipping, particularly in situations where tipping has traditionally been minimal or uncommon ~\cite{carr_how_2013}. Such an arrangement of alternatives exploits individuals' resistance to change, and thereby leading them to stick with the default suggested tip amount.\\

A particular form of digital interface that has seen a rapid rise in popularity is the AI-enhanced conversational agent, also called a chatbot. Chatbots have seen adoption in domains as diverse as financial decision-making to career counselling to psychometric testing~\cite{d2020career}~\cite{conct2022career}~\cite{muhammad2023barriers}. Fluency in language generation, using LLMs, has significantly advanced the acceptance of chatbots as a viable digital interface. In conversational choice environments like task-oriented conversational agents (\textit{i.e.}, a chatbot built for a specific task), decision-making occurs in the context of a dialogue. A user is cognitively involved in making a decision and depends on the chatbot, both for informational context as well as articulation of alternatives. In these contexts, it is trivial for the chatbot to use particular kinds of utterances to exploit the likelihood of a cognitive bias. We use `likelihood' advisedly, since all human beings are not affected by the same set of biases. Nor are they activated in the same situations. Again, we wish to reiterate that we make no value-judgement on these interactions; the chatbot could very well use utterances that diminish the likelihood of the bias, or use the bias to guide the human towards a good decision. An utterance, or a sequence of utterances, could use multiple factors such as emotional valence or dialogue complexity to activate (or diminish) the bias.\\

Based on this, we pose the following research question.

\begin{quote}

\emph{How does cognitive load, induced by the task complexity, influence susceptibility to cognitive biases in decision-making scenarios, presented by a task-oriented chatbot?}    
\end{quote}

To address this research question, we create a prototype of a task-oriented chatbot that can engage with individuals on a preference elicitation task. The chatbot is designed to present decision scenarios that are adapted from Samuelson's classic experiments with status-quo bias~\cite{samuelson_status_1988}.
First, to establish a baseline, we attempt to replicate status-quo bias using a chatbot. A simple preference elicitation task is employed for the prior dialogues. Subsequently, in our second set of experiments we increase the task complexity of the prior dialogue and observe how it influences the status-quo bias against the baseline.
We repeat the experiments with three distinct decision scenarios: budget allocation, investment decision-making, and college job selection. These scenarios have been previously replicated by Xiao et al. ~\cite{xiao_revisiting_2021}, establishing their suitability as testcases. Through these experiments, we establish a replicable relation between dialogue task complexity of a task-oriented chatbot and the manifestation of status-quo bias in subsequent decision scenarios.

The rest of this paper is structured as follows: the Background section introduces key concepts such as decision-making in conversational choice environments, status-quo bias, and cognitive load. The Related Work section reviews previous research and experiment in the intersection of cognitive biases and conversational agents, highlighting gaps in existing studies. The Hypothesis Development section presents hypotheses, focusing on the influence of task complexity on decision-making. The Methodology section explains the experimental design, including chatbot implementation, decision scenarios, and cognitive load measurement. It also describes data collection and quality control procedures. The Results section presents findings from the experiments, showing how task complexity affects cognitive load and decision-making. The Conclusion and Future Work section summarises key insights and suggests areas for further research, such as refining chatbot designs, testing different biases, and improving experimental methods.

\section{Background}
\label{sec:Background}
We begin by introducing the \textit{Conversational Choice Environment}, where task-oriented dialogue systems facilitate decision-making through structured interactions. 
Next, we examine \textit{Status Quo Bias}, a cognitive bias that leads individuals to favour existing states over change, even when better alternatives are available. We discuss foundational experiments demonstrating the presence of this bias and its implications in various decision-making contexts.
Following this, we introduce \textit{Cognitive Load}, which refers to the mental effort required to process information. We explore its impact on task performance and decision-making, particularly in chatbot interactions. We review studies on how cognitive load can be manipulated through dialogue design and how it affects user experience.


\subsection{Conversational Choice Environment}
\label{sec:Background-Conversational Choice Environment}

Task-oriented dialogue systems are designed to have algorithmic yet natural language interaction with humans. 
Primarily, the dialogue is designed to accomplish a defined set of goals or tasks, with various factors contributing to its success. A plethora of studies suggest that aspects of a chatbot dialogue such as, anthropomorphism, personality, trust, and many other factors play a key role improving human-like dialogue~\cite{chaves2021should}.


A \textit{Conversational Choice Environment} is specific instance within a dialogue facilitated by the conversational decision-maker interface technologies - such as chatbots, personal assistants, and virtual assistants - that require people to make judgements or decisions. A \textit{CCE} primarily consists of two parts: the first is the \textit{decision scenario}~\footnote{The words ``decision scenario'' and ``choice problem'' are used interchangeably in literature and in this paper.}, and the second  is the dialogue that leads up to the decision scenario which we call the \textit{decision context}~\footnote{The words ``decision context'', ``prior dialogue'', and ``prior discourse'' are also used interchangeably.}. A decision scenario comprises of a real-word or a hypothetical scenario with a finite list of alternatives. Decision scenarios naturally occur in any task-oriented dialogue such as flight booking, restaurant selection, trip planning. Implementation of the task-oriented dialogue is traditionally achieved by task-oriented dialogue systems (TODs).In recent times TODs have evolved with advancements like ANYTOD~\cite{zhao_anytod_2023} and TOD-BERT ~\cite{wu_tod-bert_2020}, enabling zero-shot learning and improved dialogue management. Large Language Models (LLMs) offer greater flexibility, as seen in SGP-TOD, achieving state-of-the-art performance~\cite{zhang_sgp-tod_2023}. The second part of a Conversational Choice Environment is decision context. Decision context can be defined as the dialogue that leads to the decision scenario. Alternatively, it is the dialogue or the discourse prior to the decision scenario. A personality test for creating an e-portfolio, or a scenario based question before suggesting an investment plan are examples of a decision context. 


A formal representation of a dialogue with decision context and choice problem is given as follows:

\[
\mathcal{D} = 
\textbf{\{}
u_1^{sys}, u_2^{usr}, u_3^{sys}, \ldots
\underbrace{ u_{t-k}^{sys}, \ldots , u_{t-1}^{usr},}_{\text{Decision context}}
\underbrace{u_t^{sys},u_{t+1}^{usr} }_{\substack{\text{Decision} \\ \text{Scenario} \\ \text{and} \\ \text{Response}}} \ldots
\textbf{\}}
\]
\label{fig:dialogue}

A dialogue \(\mathcal{D}\) consists of utterances of dialogue system (\textit{sys}) and user (\textit{usr}). The utterances between $t-k$ and $t$ act as the decision context to the decision scenario at $t$. $k-1$ is the length of the decision context.

\subsection{Status Quo Bias}
\label{sec:Background-Status Quo Bias}


Status Quo Bias is defined to be present in decision-making, when individuals favour maintaining current conditions over change, regardless of rationality. For example, in a restaurant recommendation scenario, users may prefer the chatbot’s default suggestion over exploring other options. Rational choice theory posits that decisions are made to maximise utility; however, status-quo bias contradicts this, as individuals opt for familiarity despite potential gains~\cite{masatlioglu_canonical_2014}. In decisions under uncertainty, status-quo bias manifests as a preference for familiar outcomes, even if alternative options promise superior outcomes~\cite{tversky_judgment_1974}. 
Similarly, in decisions under certainty, individuals exhibit a reluctance to deviate from existing circumstances, even when objectively advantageous alternatives exist. Thus, status-quo bias impedes rational decision-making by anchoring individuals to current states, hindering exploration of potentially beneficial alternatives. 

Over the last few decades, various experiments have shown strong evidence for status-quo bias. One of the foundational experiments for status-quo bias was conducted by  Samuelson and Zeckhauser ~\cite{samuelson_status_1988}. They performed controlled experiments using a questionnaire, containing a series of eight choice problems which they called decision scenarios. The participants in the experiments were students enrolled in economics classes at Boston University School of Management and at the Kennedy School of Government at Harvard University. A total of $486$ students took part in the study. The results of this experiments showed pronounced status-quo bias. 
In a recent experiment, Qinyu Xiao et.al ~\cite{xiao_revisiting_2021} replicated four decision scenarios from Samuelson and Zeckhauser ~\cite{samuelson_status_1988} study. They found strong empirical support for the status quo bias in three decision scenarios which includes budget allocation investment portfolios and college jobs. 

\subsection{Cognitive Load}
\label{sec:Background-Cognitive Load}

Cognitive load refers to the amount of mental effort required to process information~\cite{sweller1988cognitive}. 
Mental effort represents the portion of cognitive capacity that is actively utilised to meet the demands of a task. Consequently, it can be seen as an indicator of the actual cognitive load experienced~\cite{paas_cognitive_2003}.

Intrinsic, Extraneous, and Germane are various categories of cognitive load~\cite{paas_cognitive_2003}.
Intrinsic cognitive load arises from the inherent complexity of the information. This complexity is determined by the number of element interactions that must be understood simultaneously. Element interactions involve how different pieces of information relate to each other within a task~\cite{sweller_chapter_2011}.
Cognitive load in a piece of information can be controlled by the number of elements and interactions between the elements. When element interactions are numerous and complex, intrinsic cognitive load increases. For instance, understanding a complex mathematical equation involves multiple steps and relationships, leading to high cognitive load. Conversely, simple tasks with few element interactions generate low intrinsic cognitive load. 

Cognitive load is inversely related to task performance. Brachten et al.~\cite{brachten2020ability} demonstrated that chatbots can reduce the cognitive load needed to complete various tasks. One effective strategy to minimise cognitive load is to avoid presenting long responses or requiring users to provide complex inputs.
Similarly, Schmidhuber et al.~\cite{schmidhuber_cognitive_2021} explored the way in which the use of a chatbot affect users’ mental effort when interacting with a new software product and also to what extend does the use of a chatbot affect users’ productivity. The results showed that chatbot users experienced less cognitive load.
Fadhil~\cite{fadhil2018domain} suggested that the interactions should be concise and direct, utilising short texts or images for responses and restricted input options, such as buttons, where feasible. This approach simplifies user interactions, making it easier to process information and complete tasks efficiently. By implementing these methods, chatbots can enhance user experience and improve task performance by reducing the cognitive demands placed on users. 
Therefore, the cognitive load induced by the dialogue is dependent upon the dialogue design. By increasing the elements and interactions between the elements mental effort increases in-turn leading to higher cognitive load.

In the experimental setting the cognitive load is measured in various ways. Some common ways are, physiological, behavioural, performance, self-reporting, and subjective measure. One of the common way to measure cognitive load is recording individuals response to NASA-TLX questioner at the end of the task. NASA’s Task Load Index (TLX) scales, initially developed for aerospace applications, is a prominent subjective measure within the Cognitive Load Theory (CLT) framework. Despite its origins, the NASA-TLX remains one of the most widely recognised and established cognitive load measures. 

\section{Related Work}
\label{sec:Related Works}
The study of cognitive biases in human decision-making has been a central focus in psychology, behavioural economics, and human-computer interaction. With the increasing prevalence of conversational agents, researchers have explored their potential in both leveraging and fighting the cognitive biases~\cite{caraban_23_2019}. This section reviews previous work that has examined how conversational agents interact with individuals’ cognitive biases, targetting for behavioural influence, nudging, or decision-making interventions.

The early contributions by Ali et al. ~\cite{ali_mehenni_nudges_2021}, examined the susceptibility of children to nudging strategies employed by conversational agents and robots. Their study incorporated a modified version of the Dictator Game to investigate whether children’s generosity could be influenced by different interlocutors. The findings demonstrated that children were more likely to be influenced by conversational agents and robots than by human interlocutors, suggesting the presence of authority bias and social influence bias. These results provided early empirical evidence that chatbots and social robots could effectively exploit cognitive biases in decision-making processes, particularly among vulnerable populations.

Pilli et al. ~\cite{pilli_exploring_2024} investigated how conversational agents can serve as tools to measure cognitive biases, specifically focusing on the framing in language and loss aversion. The study employed chatbots to engage participants in structured decision-making tasks and found that framing in language and loss-aversion, the tendency to prefer avoiding losses over acquiring equivalent gains, was also observed in user responses, demonstrating that conversational agents could play a crucial role in understanding individual susceptibility to biases.

Dubiel et al. ~\cite{dubiel_impact_2024} explored the impact of voice fidelity in synthetic speech on decision-making, focusing on how auditory cues could exploit cognitive biases and heuristics. Their study investigated whether variations in voice characteristics, such as speech pace and pitch, could alter participants' choices. The results confirmed that higher-fidelity voices could influence decisions, highlighting the role of source credibility bias and affect heuristic, a mental shortcut in which people rely on their emotions to make quick decisions rather than engaging in detailed analysis. Participants were more likely to perceive and trust high-fidelity synthetic voices, raising concerns about how conversational agents could be designed to subtly nudge user decisions. This research underscored the importance of voice modulation as a persuasive tool in chatbot interactions, expanding the understanding of how non-verbal elements contribute to cognitive biases.

Building upon the works of Ali et al. ~\cite{ali_mehenni_nudges_2021}, Kalashnikova et al. ~\cite{kalashnikova_linguistic_nodate} examined linguistic nudges and their effectiveness in promoting ecological behaviour change. Their study categorized nudges into those based on reflection and those based on emotions, leveraging biases such as the status quo bias, regret aversion, and social conformity. The research found that chatbots and robots were more effective than human interlocutors in influencing participants' opinions on environmental issues. These findings reinforced previous studies by demonstrating that cognitive biases could be systematically leveraged in chatbot design to encourage behavioural shifts in various domains.

Most recently, Yamamoto ~\cite{yamamoto_suggestive_2024} introduced the concept of ``suggestive ending'' in chatbot responses. Their study built on the Ovsiankina effect, a cognitive bias where individuals feel compelled to complete unfinished tasks. By designing chatbots that leave responses open-ended or with subtle prompts, users were encouraged to engage in deeper decision-making and information-seeking behaviour. Their study found that users interacting with suggestive chatbots asked more follow-up questions and engaged in prolonged decision-making processes, indicating increased cognitive involvement. This work demonstrated that chatbots could not only influence decision outcomes but also shape the depth of user engagement with a given topic, further expanding the role of conversational agents in cognitive bias manipulation.

Ji et al. ~\cite{ji_towards_2024} in their latest study explored cognitive biases in spoken conversational search (SCS), emphasising the absence of visual cues and the complexity of unstructured dialogue. Their investigates biases such as confirmation bias, anchoring bias, and exposure effect in SCS interactions. While their work highlights the challenges of detecting cognitive biases in voice-based searches and proposes a multimodal experimental framework, it remains largely theoretical. Nonetheless, it provides a foundation for future research into mitigating biases in conversational AI systems.

These studies have explored various aspects of conversational agents like linguistic features, the affect caused by dialogue voice modulation, and length of dialogue and their role in influencing the decision-making. However, these studies focus on individual decision points or specific nudging strategies, they do not account for how previous interactions, or conversational context, influence the subsequent decision-making. Conversational agents engage users in multi-turn interactions, where the complexity of the dialogue itself imposes cognitive load that influences later decisions. In our work, we investigate how the cognitive load induced by conversational task affects subsequent decision-making processes. By examining this dimension, we aim to contribute to a deeper understanding of how conversational agents can influence cognitive biases and nudging strategies.

\section{Hypothesis Development}
\label{sec:Hypothesis Development}

The objective of this study is to examine how the cognitive load induced by the task complexity of a task-oriented chatbot’s dialogue influence individuals’ susceptibility to cognitive biases over a subsequent decision-making scenarios.

First, we investigate whether decision scenarios presented through a chatbot can replicate the findings of previous studies on status quo bias. Establishing this baseline is crucial to validating the chatbot’s effectiveness in eliciting similar behavioural patterns observed in traditional experiments. To achieve this, we design a simple task as the decision context and subsequently introduce decision scenarios that adapted from classic status-quo bias studies.

As an initial step, we hypothesise the following, which is adapted from the original hypothesis proposed in Samuelson and Zeckhauser (1988) and its subsequent replications:

\begin{enumerate}
    \item [H1] Participants presented with an alternative framed as the status-quo using chatbot will be more likely to choose that alternative than if the alternative was the non-status-quo alternative.
\end{enumerate}

Next, we examine how increasing task complexity influences individuals’ susceptibility to status quo bias in subsequent decision-making scenarios. Specifically, we replace the simple task used in the baseline experiment with a more complex task, increasing cognitive load before the decision scenario. By doing so, we aim to explore whether heightened cognitive effort makes individuals more reliant on heuristic-driven decision-making, thereby amplifying the status quo effect. This leads to our second hypothesis:

\begin{enumerate}
    \item [H2] Participants who undergo a complex task before the decision scenario will exhibit a stronger status quo effect compared to those who undergo a simple task.
\end{enumerate}

By formulating these hypotheses, we aim to systematically investigate the relationship between task complexity in chatbot interactions and status quo bias in decision-making. Establishing a baseline through H1 allows us to confirm whether chatbot-based decision scenarios align with prior studies on status quo bias. Expanding on this, H2 enables us to explore how increased cognitive load influences individuals' susceptibility to this bias.

The following section details the methodology used to test these hypotheses, including the decision scenarios, task manipulations, experimental design, and data collection procedures.

\section{Method}
\label{sec:Method}

This section outlines the methodology used to investigate how task complexity in chatbot interactions influences status quo bias in subsequent decision-making. Our approach is structured into multiple experimental conditions designed to test our hypotheses systematically.
We begin by detailing the decision scenarios, which serve as the core choice problems in our chatbot-based interactions and discusses how we carefully integrate these scenarios into our chatbot system while ensuring consistency with prior research.
Next, we describe the decision context tasks, which introduce either simple or complex preference elicitation tasks before the decision scenario. The details of the tasks are discussed in detail along with survey involved to measure necessary aspects from the tasks such as memory recall and cognitive load.
We then discuss our experimental design, including a-priori power analysis to determine an adequate sample size, data collection procedures, and quality control measures such as attention checks, seriousness screening, and open science framework registration. 
Our study is conducted as a between-subjects experiment with multiple conditions across three decision scenarios and two decision context conditions.

\subsection{Decision Scenarios}
\label{sec:Method-Decision Scenarios and Experimental Conditions}

The original study ~\cite{samuelson_status_1988}experimented with eight scenarios and five conditions. The replication study \textit{phase I} experimented two scenarios while \textit{phase II} experimented with four. 
These scenarios are experimented in different choice set configuration. For instance, budget allocation scenarios from the original study was explored under two alternatives (pairs), three alternatives (triples), and four alternatives (quads) in the choice set.
Our experiments adapted three scenarios. The three scenarios are budget allocation, investment portfolios, and college jobs. 
The rationale behind this decision is that these three scenarios have shown to have pronounced status-quo effect both in original and the replication study at least in quads configuration. 
The chatbot is designed to have have an introductory utterances like greetings which is followed by decision context and which then is followed by a decision scenario.
We took utmost care while integrating these decision scenarios into the conversational agent. The decision scenarios and the list of alternatives are are present in supplementary material.

As outlined in the original study, the neutral condition presents two alternatives without designating either as the status-quo, while in the status-quo conditions, one of the alternatives is explicitly framed as the status-quo. Following both the original and replication studies, we adopt the same three-condition design: a neutral condition, where no option is framed as the status quo, and two status quo conditions, where each alternative is separately framed as the status-quo in different trials. 

The following is one of the decision scenario and conditions.

\textbf{College Jobs Scenario Neutral Condition:}
\begin{quote}
Having just completed your graduate degree, you have two offers of teaching jobs in hand.
When evaluating teaching job offers, people typically consider the salary, the reputation of the school, the location of the school, and the likelihood of getting tenure (tenure is permanent job contract that can only be terminated for cause or under extraordinary circumstances).
Your choices are:
 \begin{itemize}
     \item College A: east coast, very prestigious school, high salary, fair chance of tenure.
     \item College B: west coast, low prestige school, high salary, good chance of tenure.
 \end{itemize}
\end{quote}

\textbf{College Jobs Scenario Status-quo Condition (College A):}

As previously mentioned, we experimented with three different decision scenarios. The alternatives of each scenarios are as follows. 

\begin{quote}
 You are currently an assistant professor at College A in the east coast. Recently, you have been approached by colleague at other university with job opportunity.
When evaluating teaching job offers, people typically consider the salary, the reputation of the school, the location of the school, and the likelihood of getting tenure (tenure is permanent job contract that can only be terminated for cause or under extraordinary circumstances).
Your choices are:
 \begin{itemize}
     \item Remain at College A: east coast, very prestigious school, high salary, fair chance of tenure.
     \item Move to College B: west coast, low prestige school, high salary, good chance of tenure.
 \end{itemize}
\end{quote}

Similarly, budget allocation decision scenarios involve allocating the National Highway Safety Commission's budget between automobile safety and highway safety programs, with different conditions. The alternatives presented vary based on a neutral condition, where both allocation options are presented equally, and status quo conditions, where the current allocation either 60\% auto safety / 40\% highway safety (60A40H) or 50\% auto safety / 50\% highway safety (50A50H) influences whether respondents choose to maintain the existing budget or shift funds between programs. 

In the investment decision making scenario, choices are presented between a moderate-risk investment (Mod. Risk/ Company A) and a high-risk investment (High Risk/ Company B). Under neutral conditions, both options are presented without any prior commitments, while in status quo conditions, the investor inherits a portfolio already allocated to either moderate-risk or high-risk investments, potentially influencing their preference to maintain the existing investment or switch. 
The details of the choice problems are made available in supplementary material.

\subsubsection{Manipulations}
\label{sec:method-decision_scenario-manipulations}

We made few adjustments to the decision scenarios to address gaps in the original study and ensure consistency across different configurations. Below, we outline the key modifications made to the phrasing, structure, and presentation of the scenarios.

Original study didn't provide phrasing for decision scenarios for pairs configuration, therefore we modified them based on the triples and quads.
In the neutral condition of the college job scenarios, as shown above, the modification made to the original scenario involves reducing the number of teaching job offers from four to two.
Similarly, in the status quo condition of the college jobs scenario, the modification made to the original scenario involves changing ``colleagues at other universities with job opportunities.'' to ``colleague at other university with job opportunity''. In all the decision scenarios, we moved from ordered list presentation of the alternatives to bullet points. 

The replication study by Xiao et al. ~\cite{xiao_revisiting_2021} included a method to assess whether participants understood the decision scenario. Participants were first shown the scenario and asked various related questions before being presented with the decision alternatives.
In our study, we adopted a different approach to maintain experimental validity. We asked the comprehension questions later in the survey but not during the conversational agents interaction. We assessed the accuracy of each individual's recall, referring to it as their Decision Scenario Recall Task Performance.   

\subsection{Prior Discourse}
\label{sec:prior-discourse}
The prior discourse incorporates two different tasks. The first task (labelled as Simple Task) adopts a preference elicitation task where five attributes are elicited from the user through closed yes or no questions. 
The design for the condition follows a conservative dialogue strategy and is not cognitively demanding.
The second task (labelled as Complex Task) uses a preference elicitation task where arithmetic comparison between the attributes and memorisation of the outcomes requires additional mental effort. Since the design of the task is cognitively demanding we hypothesise that such design will results in the increased cognitive load. 

All the condition adopts a preference elicitation task in six different domains of the Schema Guided Dialogue (SGD) dataset ~\cite{rastogi_towards_2020}, these domains are primarily properties (real estate), music, movies, calendar, banks, and messaging apps. These domains are carefully selected to have no influence on the subsequent decision scenario. This design decision is to make sure that choices made over the decision scenarios are only influenced by the affect of cognitive load due to decision context but strictly not by the domain of the task. 

The design of the Simple preference elicitation task is straight forward. Related attributes of each domain and the closed questions of each attribute are made available in supplementary material. Whereas, the complex task requires some exposition with an example as it consists of complex design structure. 

In the complex task, users are presented with a scenario where they select from various property recommendations. Similar structures are used in other domains, including artist recommendations, streaming services, calendar apps, and banking options. Each property is characterized by three attributes, such as number of bedrooms, square footage, and star rating. For the first property, real values are explicitly provided (e.g., three bedrooms, 2000 sq. ft., and a 4-star rating). However, in subsequent interactions, users must perform arithmetic calculations to determine the actual values of the attributes for other properties.
This approach follows the standard approach to increase cognitive load, which involves a combination of arithmetic reasoning and memory recall tasks~\cite{deck2015effect}. By progressively increasing the number of required calculations and memory dependencies, the cognitive effort demanded by the task also increases, making decision-making more cognitively taxing.

\newpage

\begin{enumerate}
    \item[U1.] In the following scenario choose from various property recommendations.
    \item[U2.] The first property has three bedrooms, 2000 square feet, and a 4-star rating. The second property has twice the number of bedrooms and with the same size and rating. Which one do you prefer, and why?
    \item[U3.] The third property has the same number of bedrooms as the second one but is half the size of the first one, with the same rating as the first. Which one do you prefer, and why?
    \item[U4.] The fourth property has the same number of bedrooms as the second, the same size as the third, but one less star rating than the first. Which one do you prefer, and why?
    \item [U5.] Remember the details of the fourth property. Specific information related to the property will be requested later.
\end{enumerate}

In the above example, while responding to the third dialogue utterance (U3), the user is required to put mental effort to calculate the real value of the attribute bed room as it requires user to compare the second property with the first property and deduce a real value for third property. 
By definition the interactions between the element is that which additional mental effort resulting in cognitive load. Therefore the elements here are the attributes and the interactions between these elements is the arithmetic performed by the individual. 
Refer to the supplementary material 
for other domains of  preference elicitation task.
Number of recursive calculations increases with each utterance therefore the cognitive load is expected to increase after each utterance. 
This results in the mental effort required along with progression of dialogue. 
Finally, the user is asked to memorise the deduced attribute information related to the property 
In theory, the task design must substantially increase the cognitive load of the individual. To test the cognitive load, standard NASA-TLX~\cite{pandian_nasa-tlx_2020} survey adopted for all the interactions.

\subsection{A-priori Power Analysis}
We conducted power analysis using the software programme $G*Power$~\cite{faul2009statistical} to determine the required sample size. Our objective was to achieve a power of $\beta$ of 0.80 to detect a medium effect size ($Cohen's$ $\omega$) of 0.3 (Our study investigates the difference between the effect size under various conditions rather than emphasising on establishing a statistically effect for each condition. Therefore, any where between medium to large effect size is desirable.) with a standard alpha error probability $\alpha$ of 0.05, and with 1 degree of freedom ~\cite{pancholi2009}. Based on the analysis, the target sample size for each condition was calculated to be 42 participants. Given that our study includes two decision context conditions (simple task and complex task), three decision scenarios (budget allocation, investment decision making, college jobs), and three decision scenario conditions (one neutral and two status quo conditions), we require a sample 756 participants for a between-subjects experimental design. 

\subsection{Data Collection, Quality, and Integrity}

To ensure high-quality data collection, we implemented rigorous compensation, integrity, and quality control mechanisms throughout the study.

\subsubsection{Compensation and Participant Recruitment}
Participants were recruited through Prolific, a widely used platform known for ensuring data quality and participant reliability. The estimated completion time for the survey was eight minutes, and participants were compensated according to Prolific’s recommended minimum rate of \$8 per hour, receiving more than \$0.80 after successful completion of the survey. A total of 756 participants were recruited, and measures were put in place to prevent duplicate participation. Additionally, demographic information was obtained from Prolific’s database, allowing for diversity verification and eligibility confirmation.
\subsubsection{Data Quality}
To maintain data quality participants were asked whether they had previously encountered the decision scenarios, this is to assess familiarity and ensure prior exposure did not bias their responses. 
Similarly, domain familiarity was elicited verify participants’ understanding of the subject matter and determine relation with the outcomes.
To further ensure data reliability, participants were asked to rate their seriousness at the end of the survey. If a participant indicated a lack of engagement or provided responses suggesting inattentiveness, their data was flagged for review. In cases where engagement was deemed insufficient, a data recollection process was initiated to replace unreliable responses with high-quality data. (have to provide the stats for all these)
To verify attentiveness, a memory recall task was included, where participants were asked to recall specific details from their interaction with the chatbot. 
This helped confirm whether participants had actively engaged with the task or merely responded at random.
To ensure that participants processed cognitive load naturally, they were explicitly asked not to use external aids, such as pen and paper, during cognitive tasks. This measure was necessary to maintain the intended cognitive load effects and prevent external assistance from influencing the results.
\subsubsection{Data Integrity}
An automated data integrity mechanism was also incorporated into the system. Upon submission, participant responses, captured as a Json file, were automatically emailed to both the study authors and a publicly accessible email address. This system was designed to enhance transparency and accountability, ensuring that data was securely logged and accessible for verification. The dataset will be made available for read-only public access, with access codes provided upon request.
\subsubsection{Data Cleaning and Rejection Criteria}
Participants were not automatically rejected based on a single failed attention check. Instead, a dual-layer filtering approach was applied to distinguish between minor lapses and systematic inattention. If a participant failed a Decision Context Attention Check (DCAC) but provided otherwise high-quality responses, their data was retained. However, participants who failed both attention and quality checks were removed from the dataset.
Rejected responses were replaced through recollection procedures to maintain the required sample size while ensuring the integrity of the dataset. By implementing these measures, we ensured that the final dataset was robust, reliable, and free from low-quality responses, enhancing the validity of our findings.

\subsubsection{OSF Pre-Registration}
The hypotheses and the experimental design were preregistered in the Open Science Framework (OSF) to ensure transparency and reproducibility. This preregistration includes detailed documentation of our study's objectives, methodology, decision scenarios, experimental conditions, and analysis plans. By publicly sharing our hypotheses in advance, we mitigate the risks of confirmation bias and p-hacking, ensuring that our research follows a predefined analytical framework. The preregistration also includes a-priori power analysis, justifications for sample size selection, and all materials used in the study. The complete pre-registration can be accessed at \url{https://osf.io/psxvf}, where all study materials, analysis scripts, and methodological disclosures are available for verification and replication.

\subsubsection{IRB}
The ethics of conducting research involving human participants is of utmost importance. This study was approved by the Institutional Review Board (IRB) and adheres to the ethical guidelines established by the declaration. Human Research Ethics Committee – Sciences (HREC-LS):  LS-C-25-001-Pilli-Nallur.

\subsection{Procedure}
\label{sec:procedure}
\begin{figure*}[ht]
    \centering
    \includegraphics[width=1\linewidth]{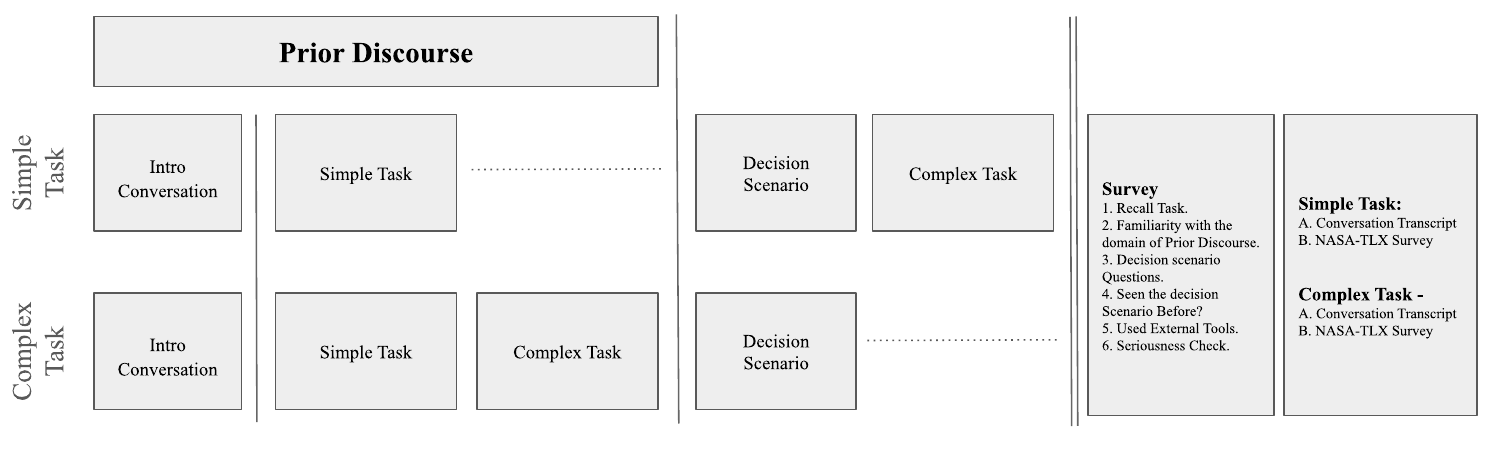}
    \caption{Chatbot interaction conditions, survey, and Cognitive load survey.}
    \label{fig:procedure}
\end{figure*}

The entire experiment was developed as an interactive Streamlit application~\cite{streamlit} and is hosted on Google Cloud Platform (GCP). The conversational user interface (CUI) is built using the Streamlit chat interface. 
On the back end, the OpenAI API powers the chatbot. Carefully crafted prompts are designed to simulate various experimental conditions. 
Additionally, the source code and condition-specific prompts are provided in the supplementary materials.

Participants began the interaction by reviewing an information sheet detailing the study’s purpose, procedures, and ethical considerations. 
Upon completion, the participants are allowed to access the experiment's homepage displaying their prolific ID and consent form.
The checkbox for consent was initially unchecked, requiring participants to actively select "Yes" to indicate their agreement before proceeding. Only after providing informed consent were participants allowed to continue to the main experiment.

The chatbot initiates the conversation with a greeting, followed by a decision context task that was assigned based on the participant’s experimental condition. 
Participants in the simple task condition engages in a straightforward preference elicitation task, while in the complex task condition undergoes a more cognitively demanding task involving arithmetic comparisons and memory recall as described in the section ~\ref{sec:prior-discourse}. Upon completion, subsequently in the same chat interaction, participants undergo a random decision scenario (discussed in previous section), where they were required to choose between a set of alternatives.

In the simple task condition, participants engage with both the simple task and the complex task; however, the complex task is introduced after the decision scenario. Towards the end of the survey, participants are first presented with the transcript of their interaction from the simple task, followed by the NASA-TLX survey to assess their cognitive load. Subsequently, they are shown the transcript of the complex task interaction, after which they complete another NASA-TLX survey. This sequential approach ensures that cognitive load is measured separately for each task, allowing for a more precise evaluation of the effect of task complexity on decision-making. Conversely, in the complex task condition, participants completed both the simple and complex tasks before being presented with the decision scenario as show in Figure ~\ref{fig:procedure}. This was a deliberate design, to make participant take complex task NASA-TLX survey in relation to simple task. This approach ensures greater accuracy compared to administering the surveys independently, minimizing recall bias and providing a more precise assessment of cognitive load.

After completing their chatbot interaction, participants were redirected to a survey page. This survey included a memory recall task to assess attentiveness, along with attention check questions to ensure data integrity. 




\section{Results}

This section presents the findings of the study, focusing on the impact of task complexity on cognitive load following the influence of task complexity on simple task and complex task. We begin with descriptive statistics, providing an overview of participant demographics and data quality assessments to ensure the validity of our dataset. Next, we analyse task complexity and cognitive load, examining how different levels of cognitive effort influenced participants' responses. Finally, we report the statistical significance and effect sizes of status-quo bias under simple and complex tasks.

\subsection{Descriptive Statistics}
\subsubsection{Demographics}
The dataset comprises 756 individuals (mean age = 41.81 years, range 18–81) who took an average of 10.20 minutes (with SD=4.82) to complete the study. The sample is slightly skewed toward female participants (53\% female, 47\% male) and is predominantly White (82.1\%), with most participants residing in either the United Kingdom (71.2\%) or the United States (27.3\%). The majority reported English as their primary language (99.7\%).
Regarding student status, 76.1\% were non-students, while 11.8\% were students, with some missing data. In terms of employment, 44.8\% were in full-time roles, 17.8\% in part-time work, and 6.4\% were unemployed and seeking jobs, while others fell into various categories, including homemakers, retirees, or those due to start a new job soon.

\subsubsection{Data Quality}
To ensure the experimental validity, several measures were implemented. These checks aimed to verify participant engagement, familiarity with the decision context, attentiveness, and authenticity of responses, minimising potential biases that could affect the results.

The familiarity with the decision context domain was generally high, with a mean score significantly above the midpoint ($t$ = 17.09, $p$ < 0.001), indicating that participants were well-acquainted with the subject matter. 
The majority of participants (698 out of 756) reported that they had not seen the decision scenarios before, while others were unsure, and the distribution was statistically significant ($p$ = 0.00). 
Additionally, none of the participants reported using external tools or expressed a need for them, confirming that decision-making relied on internal cognitive resources. 
Seriousness scores were high, with a mean score significantly above the midpoint ($t$ = 42.79, $p$ < 0.001), suggesting that participants engaged seriously with the survey tasks. 
These findings, along with Decision Context Attention Checks (DCAC), ensured that participants were attentive, engaged, and provided reliable data for hypothesis testing. 

To identify and handle outliers in response times, we used the Interquartile Range (IQR) method. This involved calculating the first quartile (Q1) and third quartile (Q3) and determining the IQR as the difference between them. Response times falling below the lower bound (Q1 minus 1.5 times the IQR) or above the upper bound (Q3 plus 1.5 times the IQR) were flagged as potential outliers. Using this method, we initially identified 53 responses for exclusion. However, some extreme outliers appeared due to technical issues, with unusually large response times (e.g., 1,329,910.313 s and 1,331,160.924 s; $n$ = 22). To ensure these were not mistakenly excluded, we cross-checked them against recorded start and completion times from Prolific and retained those that aligned with expected survey durations. Additionally, some responses had implausible negative or zero values, indicating system serious technical errors ($n$ = 5). These cases were removed excluded as well. Importantly, including or excluding these erroneous data points did not significantly impact the overall findings. To maintain a conservative approach, we excluded these unreliable responses ($n$ = 31) from the final analysis.

\subsection{Task Complexity and Cognitive Load}

To investigate the impact of task complexity on cognitive load, we analysed participants’ responses to the NASA-TLX survey after engaging with both simple and complex tasks during the chatbot's interactions. For more details please refer to the Section ~\ref{sec:procedure}. 

\begin{table}[]
\centering
\scriptsize
\begin{tabular}{@{}ccccc@{}}
\toprule
\textbf{\begin{tabular}[c]{@{}c@{}}NASA-TLX\\ Dimensions\end{tabular}} &
  \textbf{\begin{tabular}[c]{@{}c@{}}Simple Task \\ vs \\ Complex Task\end{tabular}} &
  \textbf{SC1} &
  \textbf{SC3} &
  \textbf{SC4} \\ \midrule
\multirow{2}{*}{Mental Demand}   & A    & -2.144*** & -2.232*** & -1.820*** \\ \cmidrule(l){2-5} 
                                 & B    & -2.255*** & -2.429*** & -2.570*** \\ \midrule
\multirow{2}{*}{Effort}          & A    & -1.235*** & -1.290*** & -0.978**  \\ \cmidrule(l){2-5} 
                                 & B    & -0.962**  & -1.512*** & -1.514*** \\ \midrule
\multirow{2}{*}{Performance}     & A    & -0.546*   & -0.689**  & -0.675**  \\ \cmidrule(l){2-5} 
                                 & B    & -0.389    & -1.124*** & -0.677**  \\ \midrule
\multirow{2}{*}{Frustration}     & A    & -0.956**  & -0.692**  & -0.400    \\ \cmidrule(l){2-5} 
                                 & B    & -0.750**  & -0.919**  & -1.174*** \\ \midrule
\multirow{2}{*}{Temporal Demand} & A    & -0.593*   & -0.455    & -0.439    \\ \cmidrule(l){2-5} 
                                 & B    & -0.817**  & -0.824**  & -0.489    \\ \midrule
\multirow{2}{*}{Physical Demand} & A    & -0.610*   & -0.324    & -0.233    \\ \cmidrule(l){2-5} 
                                 & B    & -0.561    & -0.429    & -0.302    \\ \bottomrule
\end{tabular}%
\caption{NASA-TLX Test Results with Effect Sizes and p-value Significance. Significance levels: *** p \textless 0.001, ** p \textless 0.01, * p \textless 0.05, + p \textless 0.10}
\label{tab:nasa-tlx-effectsize}
\end{table}

The results show that Mental Demand had the largest differences between simple and complex tasks across all scenarios (budget allocation, investment decision making, college jobs). The effect sizes, as tabulated in Table~\ref{tab:nasa-tlx-effectsize} were strong (ranging from -0.57 to -1.161), and most results were highly significant ($p$ < 0.001, $p$ < 0.01, or $p$ < 0.05). This suggests that participants experienced much higher mental workload when performing complex tasks.
\begin{figure}[ht]
    \centering
    \includegraphics[width=1.0\linewidth]{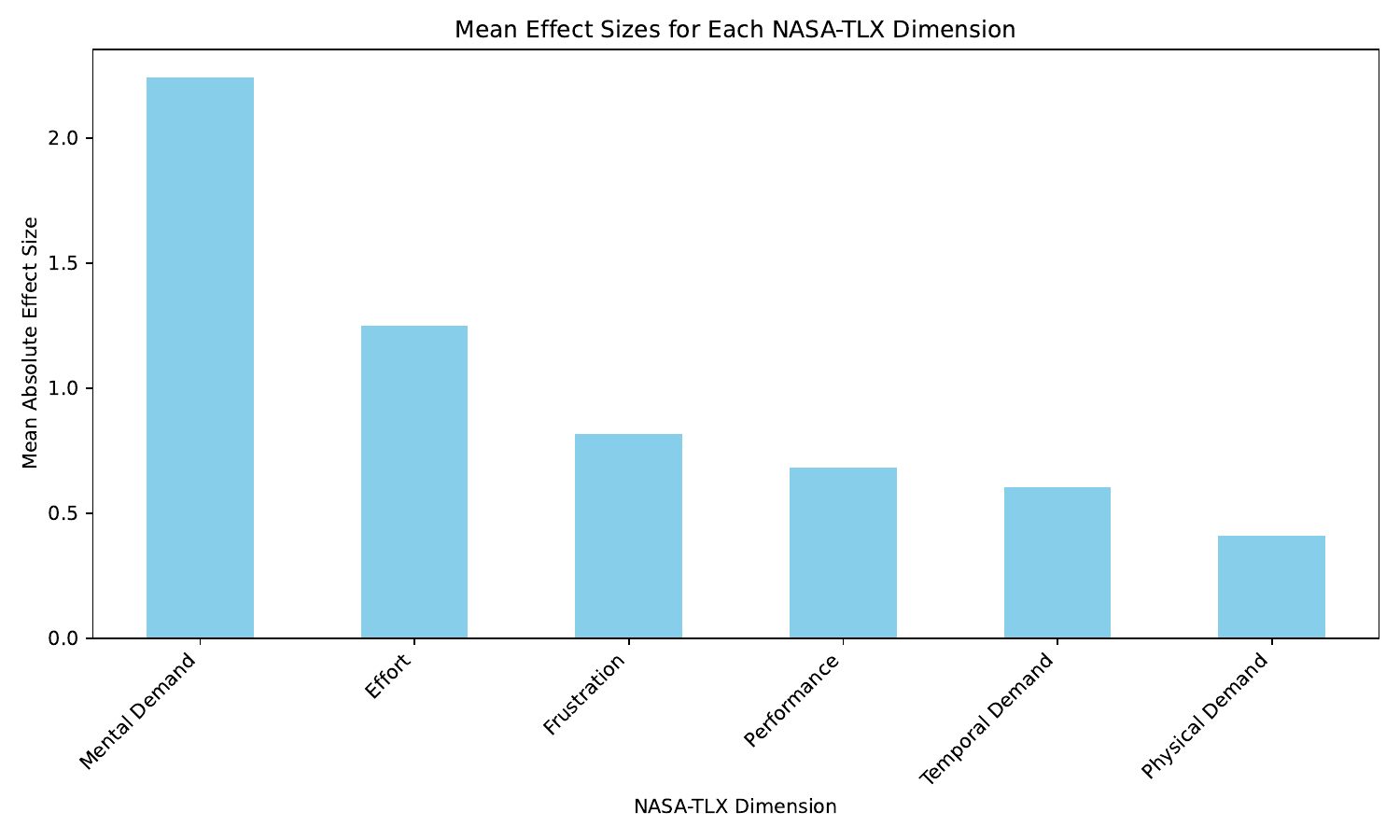}
    \caption{Mean Effect Sizes for Each NASA-TLX Dimension}
    \label{fig:ntlx-effect-sizes}
\end{figure}
Effort also showed significant differences in multiple scenarios, especially in budget allocation and investment decision making, with effect sizes up to -0.864 and p-values below 0.001 as shown in Figure ~\ref{fig:ntlx-effect-sizes}. This indicates that participants had to put in more effort for complex tasks.
For Performance, Frustration, and Temporal Demand, some significant effects were found, but the effect sizes were smaller compared to Mental Demand and Effort. Physical Demand showed the least significant differences, suggesting that physical effort was not a major factor in these tasks.
Overall, Mental Demand, as shown in the Figure ~\ref{fig:ntlx-effect-sizes} was the most affected by task complexity, making it the most important factor.

\subsection{Task Complexity and Status Quo Bias}

\begin{table*}[ht]
        \centering
        \scriptsize
        \renewcommand{\arraystretch}{1.5} 
        \label{sec:sqb framing vs non-sqb framing}%
        \begin{tabular}{lllllllll}%
          \hline%
          \multirow{3}{*}{Alternatives}&\multicolumn{3}{c}{Choice rates}&&\multicolumn{4}{c}{Status quo framing vs. non-status quo framing}\\%
          \cline{2%
            -%
            4}%
          \cline{6%
            -%
            9}%
                                                                  &   &   &   &   &   &   &   &   \\%
          &SQ&N&NSQ&&$\chi^2$&\textit{p}&Odds ratio \newline%
          (95\% CI)&Cohen's h \newline%
          (95\%CI)\\%
          \hline%
          \textbf{Scenario 1: Budget allocation ratios} &   &   &   &   &   &   &   &   \\%
          60A40H                                                   & 16/41 (0.39) & 12/42 (0.29) & 4/42 (0.1) &  & 9.87& 0.002*& 6.08 [1.82, 20.31]& 0.72 [-0.13, 1.58] \\%
          50A50H                                                   & 38/42 (0.9) & 30/42 (0.71) & 25/41 (0.61) &  &  &  &  &  \\%
          \textbf{Scenario 2: Investment portfolios}    &   &   &   &   &   &   &   &   \\%
          Mod. Risk                                                & 32/42 (0.76) & 32/40 (0.8) & 33/42 (0.79) &  & 0.07& 0.794& 0.87 [0.31, 2.43]& -0.06 [-0.78, 0.67] \\%
          High Risk                                                & 9/42 (0.21) & 8/40 (0.2) & 10/42 (0.24) &  &  &  &  &   \\%
          \textbf{Scenario 3: College jobs }            &   &   &   &   &   &   &   &   \\%
          College A                                                & 25/41 (0.61) & 19/41 (0.46) & 21/41 (0.51) &  & 0.79& 0.373& 1.49 [0.62, 3.58]& 0.2 [-0.42, 0.82] \\%
          College B                                                & 20/41 (0.49) & 22/41 (0.54) & 16/41 (0.39) &  &  &  &  &  \\%
          \hline%
        \end{tabular}

        
        \caption{Simple Task - Status quo framing vs. non-status quo framing}
        \label{tab:ST-SQNSQ}
      \end{table*}

\begin{table*}[ht]
        \centering
        \scriptsize
        \renewcommand{\arraystretch}{1.5} 
        \label{sec:sqb framing vs non-sqb framing}%
        \begin{tabular}{lllllllll}%
          \hline%
          \multirow{3}{*}{Alternatives}&\multicolumn{3}{c}{Choice rates}&&\multicolumn{4}{c}{Status quo framing vs. non-status quo framing}\\%
          \cline{2%
            -%
            4}%
          \cline{6%
            -%
            9}%
                                                                  &   &   &   &   &   &   &   &   \\%
          &SQ&N&NSQ&&$\chi^2$&\textit{p}&Odds ratio \newline%
          (95\% CI)&Cohen's h \newline%
          (95\%CI)\\%
          \hline%
          \textbf{Scenario 1: Budget allocation ratios} &   &   &   &   &   &   &   &   \\%
          60A40H                                                   & 20/37 (0.54) & 9/39 (0.23) & 7/41 (0.17) &  & 11.75& 0.001*& 5.71 [2.02, 16.15]& 0.8 [0.07, 1.53] \\%
          50A50H                                                   & 34/41 (0.83) & 30/39 (0.77) & 17/37 (0.46) &  &  &  &  &  \\%
          \textbf{Scenario 2: Investment portfolios}    &   &   &   &   &   &   &   &   \\%
          Mod. Risk                                                & 33/39 (0.85) & 35/41 (0.85) & 27/36 (0.75) &  & 1.08& 0.298& 1.83 [0.58, 5.8]& 0.24 [-0.57, 1.06] \\%
          High Risk                                                & 9/36 (0.25) & 6/41 (0.15) & 6/39 (0.15) &  &  &  &  &   \\%
          \textbf{Scenario 3: College jobs }            &   &   &   &   &   &   &   &   \\%
          College A                                                & 24/40 (0.6) & 20/39 (0.51) & 19/40 (0.48) &  & 1.26& 0.262& 1.66 [0.68, 4.02]& 0.25 [-0.38, 0.88] \\%
          College B                                                & 21/40 (0.52) & 19/39 (0.49) & 16/40 (0.4) &  &  &  &  &  \\%
          \hline%
        \end{tabular}
        \newline
        \caption{Complex Task - Status quo framing vs. non-status quo framing}
        \label{tab:CT-SQNSQ}
      \end{table*}

Our study investigated the influence of task complexity on bias detection across three decision-making contexts: budget allocation (BA), investment portfolio (IDM), and college jobs (CJ). We compared our findings to the original study (Samuelson\cite{samuelson_status_1988}).

\begin{figure*}[ht]
    \centering
    \includegraphics[width=1.0\linewidth]{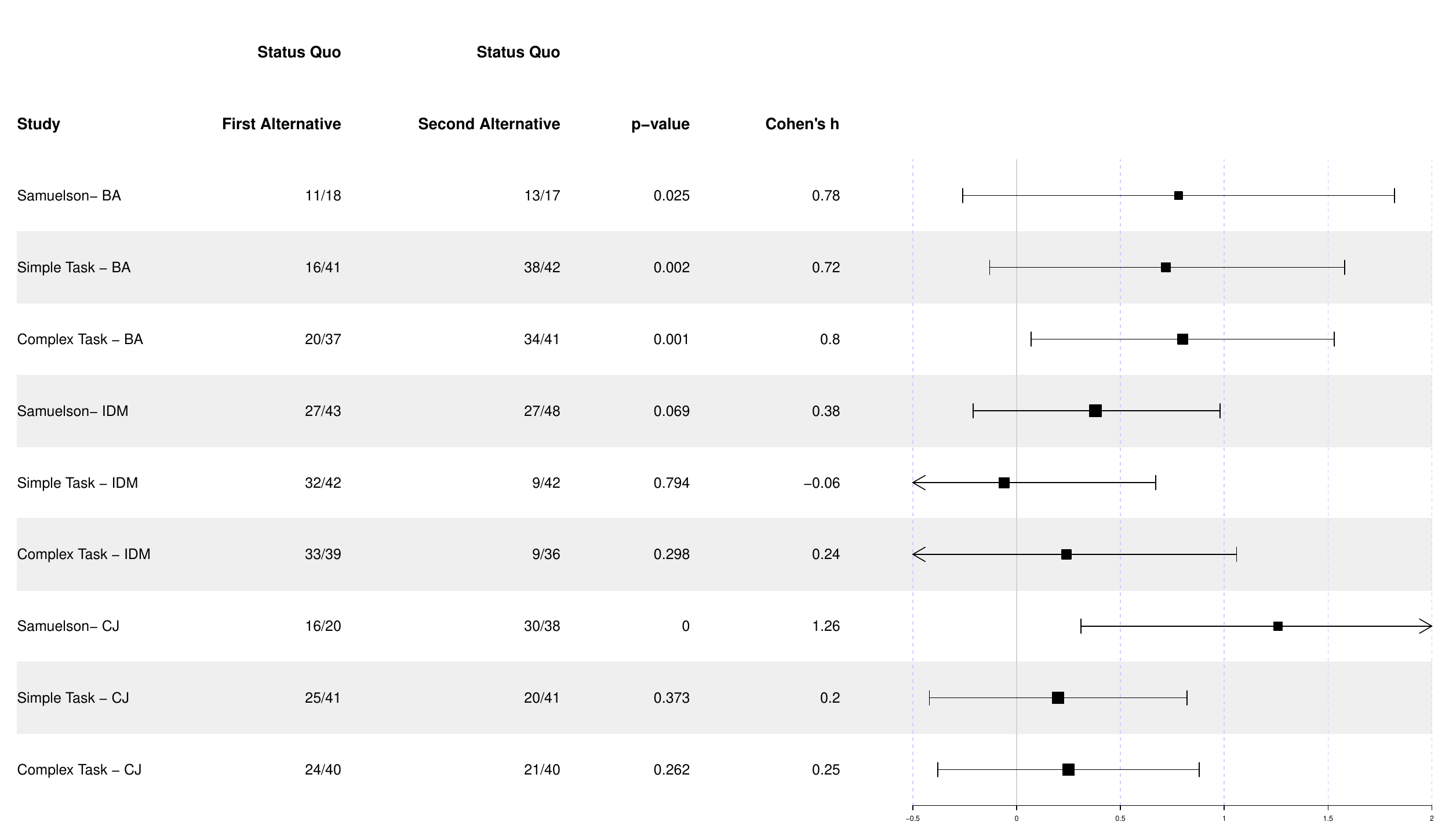}
    \caption{Comparison of Effect-sizes across various studies. Size of the block indicates the size of the sample.}
    \label{fig:forest-plot}
\end{figure*}

\subsubsection{Budget Allocation Scenario}

In the Samuelson study, when 60A40H was the status quo alternative, 61\% (11/18) of participants chose it. However, when 50A50H was framed as the status quo, 76\% (13/17) of participants preferred it. This difference was statistically significant ($p$ = 0.025), with a moderate to large effect size ($Cohen’s$ $h$  = 0.78). However, the confidence interval ($CI:$ -0.26 to 1.82) was wide, suggesting some uncertainty about the true effect size.

Similarly, in our simple task condition, 39\% (16/41) of participants preferred 60A40H, while 90\% (38/42) chose 50A50H when it was the status quo. This difference was statistically significant ($p$ = 0.002), with an effect size of $h$  = 0.72 as show in the Table ~\ref{tab:ST-SQNSQ}. While the confidence interval ($CI:$ -0.13 to 1.58) remained wide, the effect size closely matched that of the Samuelson's study, confirming that status quo bias was successfully detected in this condition.

In the complex task condition, 54\% (20/37) of participants preferred 60A40H, while 83\% (34/41) selected 50A50H as the status quo alternative as show in the Table ~\ref{tab:CT-SQNSQ}. This difference was statistically significant ($p$ = 0.001), with the largest effect size ($h$  = 0.80) observed in our study. Notably, the confidence interval ($CI:$ 0.07 to 1.53) did not include zero, providing stronger evidence that higher task complexity increases susceptibility to status quo bias.

The effect sizes in our study closely aligned with those in the Samuelson study, confirming the presence of status quo bias, particularly in the simple task condition. However, while the confidence intervals in both the Samuelson and simple task conditions included zero, indicating some uncertainty, the complex task condition demonstrated the strongest effect size with a confidence interval that excluded zero, providing some evidence that higher task complexity influences status-quo bias.

\subsubsection{Investment Portfolio Scenario}

In the Samuelson IDM condition, 62.8\% (27/43) of participants chose the moderate-risk alternative when framed as the status quo, while 56.3\% (27/48) preferred the high-risk alternative when it was the status quo alternative. The effect size was small to moderate ($h$ = 0.38), and although the p-value = 0.069 indicated a trend toward significance, the confidence interval ($CI:$ -0.21 to 0.98) suggests some variability in the effect.

In our simple task condition, 76.2\% (32/42) of participants chose the moderate-risk alternative when it was the status quo, whereas 21.4\% (9/42) selected the high-risk alternative when it was framed as the status-quo alternative. However, the effect size was negligible ($h$ = -0.06), and the p-value = 0.794, indicating no significant difference between conditions. The confidence interval ($CI:$ -0.78 to 0.67) was wide, suggesting a high level of uncertainty in the effect.

In contrast, the complex task condition showed a positive shift toward the baseline effect observed in the Samuelson study. Here, 84.6\% (33/39) of participants chose the moderate-risk alternative as the status quo, while 25.0\% (9/36) selected the high-risk alternative when framed as the default. The effect size was small but larger than in the simple task condition ($h$  = 0.24), suggesting a trend toward increased bias under higher task complexity. Although the p-value = 0.298 did not reach statistical significance, the effect size's movement closer to the Samuelson's baseline ($h$  = 0.38) as shown in the forest plot Figure ~\ref{fig:forest-plot}. This provides some evidence that increased task complexity may strengthen status-quo bias, supporting our hypothesis to a certain extent.

The effect sizes in our study suggest a directional trend toward status quo bias in investment decision-making, with the complex task condition moving closer to the baseline effect observed in the Samuelson study. While the confidence intervals in both the simple and complex task conditions remained wide and included zero, indicating uncertainty, the increasing effect size under higher task complexity provides some evidence that cognitive load may enhance status-quo bias in investment choices.

\subsubsection{College Jobs Scenario}
The Samuelson study demonstrated a large and highly significant status-quo bias effect, with 80.0\% (16/20) of participants selecting the East Coast (College A) alternative when framed as the status quo, and 78.9\% (30/38) choosing the West Coast (College B) alternative when it was in the status-quo position. The effect was highly significant ($p$ = 0.000, $h$  = 1.26), and the confidence interval ($CI:$ 0.31 to 2.21) confirmed a strong and reliable bias effect, reinforcing the tendency to prefer the status-quo option.

In our Simple Task condition, 61.0\% (25/41) of participants chose the East Coast alternative when presented as the status quo, while 48.8\% (20/41) preferred the West Coast alternative when framed as status-quo alternative. The effect size was small ($h$  = 0.2), and the result was not statistically significant ($p$ = 0.373). The confidence interval ($CI:$ -0.42 to 0.82) was broad, indicating variability in the observed effect. 

In the Complex Task condition, a negligible but directional trend towards hypothesis was observed. 60.0\% (24/40) of participants chose the East Coast alternative when it was the status quo, while 52.5\% (21/40) selected the West Coast alternative when framed as the default. The effect size ($h$  = 0.25) was slightly higher than in the Simple Task condition, suggesting a minor increase in bias under higher task complexity. However, the p-value ($p$ = 0.262) did not reach statistical significance, and the confidence interval ($CI:$ -0.38 to 0.88) remained wide, indicating continued uncertainty about the effect.

The effect sizes in our study suggest a much weaker status quo bias in the college jobs scenario compared to the Samuelson study, where the effect was large and highly significant. In both the simple and complex task conditions, the effect sizes were small, and the confidence intervals remained wide and included zero, indicating substantial uncertainty. While the complex task condition showed a slight increase in effect size compared to the simple task, the absence of statistical significance suggests that task complexity did not meaningfully amplify status quo bias in this decision context.

One possible explanation for this outcome is the demographic composition of our sample, as the majority of participants were from the UK, while the decision scenario was framed for U.S. geography. Given that participants may not have had strong preferences or familiarity with the locations, could have caused this. 

\subsection{Response Times and Recall Task Performances}

We further investigated the average response times and recall task performance of the decision scenarios. Participants in the complex task took significantly longer to respond compared to the simple task, reflecting increased cognitive effort. Despite this, their recall performance remained high and unchanged across conditions, indicating full engagement with the decision scenarios regardless of cognitive load.

The average response time for the simple task was 53.58s, while participants in the complex task took significantly longer which is 76.24s. Across all decision scenarios, response times in the complex task were consistently higher than in the simple task (\( t \) = 5.504, \( p < 0.001 \)). These response times also aligned with cognitive load survey, indicating that participants experienced greater cognitive effort under increased task complexity. 
Despite the longer response time and higher cognitive load, decision scenario recall task performance remained high in both conditions, with no significant difference between the simple task (\( M = 0.875, SD = 0.069 \)) and the complex task (\( M = 0.858, SD = 0.063 \)). This suggests that participants were fully engaged with the decision scenarios regardless of cognitive load, though they required more time to complete decisions under higher complexity.

\subsection{Inherent Bias in Decision Scenario Alternatives}
In the budget allocation scenario, there was a clear preference for the 50A50H alternative, even when 60A40H was set as the status quo. Specifically, 61\% of participants still preferred 50A50H, indicating an inherent bias toward equal budget distribution as shown in Table ~\ref{tab:ST-SQNSQ}. However, this preference diminished under the complex task, where the difference between choices narrowed to 46\%. The neutral condition revealed the strongest bias, with a 29\% vs. 71\% split, further highlighting the natural inclination toward 50A50H (Refer Table ~\ref{tab:CT-SQNSQ}). Importantly, in the complex task condition, this inherent bias remained relatively stable (23\% vs. 77\%), suggesting that increasing task complexity did not significantly alter this trend.

In the investment decision-making scenario, participants showed a strong bias toward Moderate Risk investments. When High Risk was set as the status quo, 75\% still preferred Moderate Risk, indicating a strong tendency to avoid higher-risk choices. This bias was even more pronounced in the neutral condition, where over 80\% of participants chose Moderate Risk in both the simple and complex task conditions. Moreover, increasing task complexity did not significantly alter this preference, suggesting that the bias toward Moderate Risk investments is robust and largely unaffected by task difficulty. In contrast, we did not observe a similar inherent bias in the college jobs scenario. Preferences between College A and College B remained more balanced across conditions, with no strong tendency toward one alternative.

\section{Conclusion}
This study examined the influence of prior discourse complexity in conversational agents on subsequent decision-making, particularly its influence on status quo bias. Across all decision scenarios, we observed a common trend: while increasing task complexity consistently shifted the effect size towards our hypothesis, the change was not statistically significant in most cases. However, in the budget allocation scenario, the complex task condition resulted in a reliable detection of status quo bias, as its confidence intervals did not include zero. This suggests that cognitive load can meaningfully reinforce biases under specific conditions. In the investment decision-making scenario, the shift from no effect in the simple task to a small-to-medium effect size in the complex task highlights the need for further exploration. While the effect did not reach statistical significance, the consistent directional movement suggests that heightened cognitive load may play a role in exacerbating status-quo bias.

These findings have implications for digital nudging and behavioural economics. Conversational agents are increasingly used to guide users in various decision-making tasks, and our results suggest that dialogue complexity may shape the way biases manifest. If cognitive load strengthens biases rather than mitigating them, designers of digital nudges must be cautious in structuring interactions to avoid unintended manipulations.

Our findings highlight the presence of inherent biases in decision scenarios, where certain alternatives naturally attract preference regardless of status quo framing. For example, in the budget allocation scenario, participants exhibited a strong preference for equal distribution (50A50H), even when 60A40H was framed as the default. Similarly, in the investment decision-making scenario, moderate-risk options were favoured regardless of framing. These inherent biases can impact experimental outcomes by masking or exaggerating the effects, making it crucial for the future experiments on these decision scenarios carefully design experiments that take baseline preferences into account. 

Future research should investigate the threshold at which cognitive load meaningfully impacts bias expression and explore additional cognitive biases such as framing, anchoring, and loss-aversion. More rigorous experimental designs, including larger sample sizes, could provide further insights into the interaction between prior dialogue complexity and decision-making. 

\begin{acks}
This work was conducted with the financial support of the Research Ireland Centre for Research Training in Digitally-Enhanced Reality (d-real) under Grant No. 18/CRT/6224. For the purpose of Open Access, the author has applied a CC BY public copyright licence to any Author Accepted Manuscript version arising from this submission
\end{acks}

\bibliographystyle{ACM-Reference-Format}
\bibliography{bibliography}


\begin{thebibliography}{36}


\ifx \showCODEN    \undefined \def \showCODEN     #1{\unskip}     \fi
\ifx \showISBNx    \undefined \def \showISBNx     #1{\unskip}     \fi
\ifx \showISBNxiii \undefined \def \showISBNxiii  #1{\unskip}     \fi
\ifx \showISSN     \undefined \def \showISSN      #1{\unskip}     \fi
\ifx \showLCCN     \undefined \def \showLCCN      #1{\unskip}     \fi
\ifx \shownote     \undefined \def \shownote      #1{#1}          \fi
\ifx \showarticletitle \undefined \def \showarticletitle #1{#1}   \fi
\ifx \showURL      \undefined \def \showURL       {\relax}        \fi
\providecommand\bibfield[2]{#2}
\providecommand\bibinfo[2]{#2}
\providecommand\natexlab[1]{#1}
\providecommand\showeprint[2][]{arXiv:#2}

\bibitem[Ali~Mehenni et~al\mbox{.}(2021)]%
        {ali_mehenni_nudges_2021}
\bibfield{author}{\bibinfo{person}{Hugues Ali~Mehenni}, \bibinfo{person}{Sofiya Kobylyanskaya}, \bibinfo{person}{Ioana Vasilescu}, {and} \bibinfo{person}{Laurence Devillers}.} \bibinfo{year}{2021}\natexlab{}.
\newblock \showarticletitle{Nudges with {Conversational} {Agents} and {Social} {Robots}: {A} {First} {Experiment} with {Children} at a {Primary} {School}}.
\newblock In \bibinfo{booktitle}{\emph{Conversational {Dialogue} {Systems} for the {Next} {Decade}}}, \bibfield{editor}{\bibinfo{person}{Luis~Fernando D'Haro}, \bibinfo{person}{Zoraida Callejas}, {and} \bibinfo{person}{Satoshi Nakamura}} (Eds.). \bibinfo{publisher}{Springer}, \bibinfo{address}{Singapore}, \bibinfo{pages}{257--270}.
\newblock
\showISBNx{9789811583957}
\href{https://doi.org/10.1007/978-981-15-8395-7_19}{doi:\nolinkurl{10.1007/978-981-15-8395-7_19}}


\bibitem[Brachten et~al\mbox{.}(2020)]%
        {brachten2020ability}
\bibfield{author}{\bibinfo{person}{Florian Brachten}, \bibinfo{person}{Felix Br{\"u}nker}, \bibinfo{person}{Nicholas~RJ Frick}, \bibinfo{person}{Bj{\"o}rn Ross}, {and} \bibinfo{person}{Stefan Stieglitz}.} \bibinfo{year}{2020}\natexlab{}.
\newblock \showarticletitle{On the ability of virtual agents to decrease cognitive load: an experimental study}.
\newblock \bibinfo{journal}{\emph{Information Systems and e-Business Management}} \bibinfo{volume}{18}, \bibinfo{number}{2} (\bibinfo{year}{2020}), \bibinfo{pages}{187--207}.
\newblock


\bibitem[Caraban et~al\mbox{.}(2019)]%
        {caraban_23_2019}
\bibfield{author}{\bibinfo{person}{Ana Caraban}, \bibinfo{person}{Evangelos Karapanos}, \bibinfo{person}{Daniel Gonçalves}, {and} \bibinfo{person}{Pedro Campos}.} \bibinfo{year}{2019}\natexlab{}.
\newblock \showarticletitle{23 {Ways} to {Nudge}: {A} {Review} of {Technology}-{Mediated} {Nudging} in {Human}-{Computer} {Interaction}}. In \bibinfo{booktitle}{\emph{Proceedings of the 2019 {CHI} {Conference} on {Human} {Factors} in {Computing} {Systems}}} \emph{(\bibinfo{series}{{CHI} '19})}. \bibinfo{publisher}{Association for Computing Machinery}, \bibinfo{address}{New York, NY, USA}, \bibinfo{pages}{1--15}.
\newblock
\showISBNx{978-1-4503-5970-2}
\href{https://doi.org/10.1145/3290605.3300733}{doi:\nolinkurl{10.1145/3290605.3300733}}


\bibitem[Carr(2013)]%
        {carr_how_2013}
\bibfield{author}{\bibinfo{person}{Austin Carr}.} \bibinfo{year}{2013}\natexlab{}.
\newblock \bibinfo{title}{How {Square} {Register}’s {UI} {Guilts} {You} {Into} {Leaving} {Tips}}.
\newblock
\urldef\tempurl%
\url{https://www.fastcompany.com/3022182/how-square-registers-ui-guilts-you-into-leaving-tips}
\showURL{%
\tempurl}


\bibitem[Carter et~al\mbox{.}(2007)]%
        {carter2007behavioral}
\bibfield{author}{\bibinfo{person}{Craig~R Carter}, \bibinfo{person}{Lutz Kaufmann}, {and} \bibinfo{person}{Alex Michel}.} \bibinfo{year}{2007}\natexlab{}.
\newblock \showarticletitle{Behavioral supply management: a taxonomy of judgment and decision-making biases}.
\newblock \bibinfo{journal}{\emph{International Journal of Physical Distribution \& Logistics Management}} \bibinfo{volume}{37}, \bibinfo{number}{8} (\bibinfo{year}{2007}), \bibinfo{pages}{631--669}.
\newblock


\bibitem[Chaves and Gerosa(2021)]%
        {chaves2021should}
\bibfield{author}{\bibinfo{person}{Ana~Paula Chaves} {and} \bibinfo{person}{Marco~Aurelio Gerosa}.} \bibinfo{year}{2021}\natexlab{}.
\newblock \showarticletitle{How should my chatbot interact? A survey on social characteristics in human--chatbot interaction design}.
\newblock \bibinfo{journal}{\emph{International Journal of Human--Computer Interaction}} \bibinfo{volume}{37}, \bibinfo{number}{8} (\bibinfo{year}{2021}), \bibinfo{pages}{729--758}.
\newblock


\bibitem[Con{\c{t}} et~al\mbox{.}(2022)]%
        {conct2022career}
\bibfield{author}{\bibinfo{person}{Miruna Con{\c{t}}}, \bibinfo{person}{Aurelia Ciupe}, \bibinfo{person}{Bogdan Orza}, \bibinfo{person}{Irina Cohu{\c{t}}}, {and} \bibinfo{person}{Georgiana Ni{\c{t}}u}.} \bibinfo{year}{2022}\natexlab{}.
\newblock \showarticletitle{Career counseling chatbot using microsoft bot frameworks}. In \bibinfo{booktitle}{\emph{2022 IEEE Global Engineering Education Conference (EDUCON)}}. IEEE, \bibinfo{pages}{1387--1392}.
\newblock


\bibitem[Deck and Jahedi(2015)]%
        {deck2015effect}
\bibfield{author}{\bibinfo{person}{Cary Deck} {and} \bibinfo{person}{Salar Jahedi}.} \bibinfo{year}{2015}\natexlab{}.
\newblock \showarticletitle{The effect of cognitive load on economic decision making: A survey and new experiments}.
\newblock \bibinfo{journal}{\emph{European Economic Review}}  \bibinfo{volume}{78} (\bibinfo{year}{2015}), \bibinfo{pages}{97--119}.
\newblock


\bibitem[Dubiel et~al\mbox{.}(2024)]%
        {dubiel_impact_2024}
\bibfield{author}{\bibinfo{person}{Mateusz Dubiel}, \bibinfo{person}{Anastasia Sergeeva}, {and} \bibinfo{person}{Luis~A. Leiva}.} \bibinfo{year}{2024}\natexlab{}.
\newblock \showarticletitle{Impact of {Voice} {Fidelity} on {Decision} {Making}: {A} {Potential} {Dark} {Pattern}?}. In \bibinfo{booktitle}{\emph{Proceedings of the 29th {International} {Conference} on {Intelligent} {User} {Interfaces}}}. \bibinfo{publisher}{ACM}, \bibinfo{address}{Greenville SC USA}, \bibinfo{pages}{181--194}.
\newblock
\showISBNx{979-8-4007-0508-3}
\href{https://doi.org/10.1145/3640543.3645202}{doi:\nolinkurl{10.1145/3640543.3645202}}


\bibitem[D’Silva et~al\mbox{.}(2020)]%
        {d2020career}
\bibfield{author}{\bibinfo{person}{Godson D’Silva}, \bibinfo{person}{Megh Jani}, \bibinfo{person}{Vipul Jadhav}, \bibinfo{person}{Amit Bhoir}, {and} \bibinfo{person}{Prithvi Amin}.} \bibinfo{year}{2020}\natexlab{}.
\newblock \showarticletitle{Career counselling chatbot using cognitive science and artificial intelligence}. In \bibinfo{booktitle}{\emph{Advanced Computing Technologies and Applications: Proceedings of 2nd International Conference on Advanced Computing Technologies and Applications—ICACTA 2020}}. Springer, \bibinfo{pages}{1--9}.
\newblock


\bibitem[Eidelman and Crandall(2012)]%
        {eidelman_bias_2012}
\bibfield{author}{\bibinfo{person}{Scott Eidelman} {and} \bibinfo{person}{Christian~S. Crandall}.} \bibinfo{year}{2012}\natexlab{}.
\newblock \showarticletitle{Bias in {Favor} of the {Status} {Quo}}.
\newblock \bibinfo{journal}{\emph{Social and Personality Psychology Compass}} \bibinfo{volume}{6}, \bibinfo{number}{3} (\bibinfo{year}{2012}), \bibinfo{pages}{270--281}.
\newblock
\showISSN{1751-9004}
\href{https://doi.org/10.1111/j.1751-9004.2012.00427.x}{doi:\nolinkurl{10.1111/j.1751-9004.2012.00427.x}}
\newblock
\shownote{\_eprint: https://onlinelibrary.wiley.com/doi/pdf/10.1111/j.1751-9004.2012.00427.x}.


\bibitem[Fadhil(2018)]%
        {fadhil2018domain}
\bibfield{author}{\bibinfo{person}{Ahmed Fadhil}.} \bibinfo{year}{2018}\natexlab{}.
\newblock \showarticletitle{Domain specific design patterns: designing for conversational user interfaces}.
\newblock \bibinfo{journal}{\emph{arXiv preprint arXiv:1802.09055}} (\bibinfo{year}{2018}).
\newblock


\bibitem[Faul et~al\mbox{.}(2009)]%
        {faul2009statistical}
\bibfield{author}{\bibinfo{person}{Franz Faul}, \bibinfo{person}{Edgar Erdfelder}, \bibinfo{person}{Axel Buchner}, {and} \bibinfo{person}{Albert-Georg Lang}.} \bibinfo{year}{2009}\natexlab{}.
\newblock \showarticletitle{Statistical power analyses using G* Power 3.1: Tests for correlation and regression analyses}.
\newblock \bibinfo{journal}{\emph{Behavior research methods}} \bibinfo{volume}{41}, \bibinfo{number}{4} (\bibinfo{year}{2009}), \bibinfo{pages}{1149--1160}.
\newblock


\bibitem[Inc.(2019)]%
        {streamlit}
\bibfield{author}{\bibinfo{person}{Streamlit Inc.}} \bibinfo{year}{2019}\natexlab{}.
\newblock \bibinfo{title}{Streamlit: A Faster Way to Build and Share Data Apps}.
\newblock \bibinfo{howpublished}{Available at \url{https://streamlit.io/}}.
\newblock
\newblock
\shownote{Accessed: 2025-02-27}.


\bibitem[Ji et~al\mbox{.}(2024)]%
        {ji_towards_2024}
\bibfield{author}{\bibinfo{person}{Kaixin Ji}, \bibinfo{person}{Sachin Pathiyan~Cherumanal}, \bibinfo{person}{Johanne~R. Trippas}, \bibinfo{person}{Danula Hettiachchi}, \bibinfo{person}{Flora~D. Salim}, \bibinfo{person}{Falk Scholer}, {and} \bibinfo{person}{Damiano Spina}.} \bibinfo{year}{2024}\natexlab{}.
\newblock \showarticletitle{Towards {Detecting} and {Mitigating} {Cognitive} {Bias} in {Spoken} {Conversational} {Search}}. In \bibinfo{booktitle}{\emph{26th {International} {Conference} on {Mobile} {Human}-{Computer} {Interaction}}}. \bibinfo{publisher}{ACM}, \bibinfo{address}{Melbourne VIC Australia}, \bibinfo{pages}{1--10}.
\newblock
\showISBNx{979-8-4007-0506-9}
\href{https://doi.org/10.1145/3640471.3680245}{doi:\nolinkurl{10.1145/3640471.3680245}}


\bibitem[Kahneman(2011)]%
        {kahneman_thinking_2011}
\bibfield{author}{\bibinfo{person}{Daniel Kahneman}.} \bibinfo{year}{2011}\natexlab{}.
\newblock \bibinfo{booktitle}{\emph{Thinking, fast and slow}}.
\newblock \bibinfo{publisher}{Farrar, Straus and Giroux}, \bibinfo{address}{New York, NY, US}.
\newblock
\showISBNx{978-0-374-27563-1 978-1-4299-6935-2}
\newblock
\shownote{Pages: 499}.


\bibitem[Kalashnikova et~al\mbox{.}({[n.\,d.]})]%
        {kalashnikova_linguistic_nodate}
\bibfield{author}{\bibinfo{person}{Natalia Kalashnikova}, \bibinfo{person}{Ioana Vasilescu}, {and} \bibinfo{person}{Laurence Devillers}.} \bibinfo{year}{[n.\,d.]}\natexlab{}.
\newblock \showarticletitle{Linguistic {Nudges} and {Verbal} {Interaction} with {Robots}, {Smart}-{Speakers}, and {Humans}}.
\newblock  (\bibinfo{year}{[n.\,d.]}).
\newblock


\bibitem[Masatlioglu and Ok(2014)]%
        {masatlioglu_canonical_2014}
\bibfield{author}{\bibinfo{person}{Yusufcan Masatlioglu} {and} \bibinfo{person}{Efe~A. Ok}.} \bibinfo{year}{2014}\natexlab{}.
\newblock \showarticletitle{A {Canonical} {Model} of {Choice} with {Initial} {Endowments}}.
\newblock \bibinfo{journal}{\emph{The Review of Economic Studies}} \bibinfo{volume}{81}, \bibinfo{number}{2} (\bibinfo{date}{April} \bibinfo{year}{2014}), \bibinfo{pages}{851--883}.
\newblock
\showISSN{0034-6527}
\href{https://doi.org/10.1093/restud/rdt037}{doi:\nolinkurl{10.1093/restud/rdt037}}


\bibitem[Muhammad(2023)]%
        {muhammad2023barriers}
\bibfield{author}{\bibinfo{person}{Rifqi Muhammad}.} \bibinfo{year}{2023}\natexlab{}.
\newblock \showarticletitle{Barriers and effectiveness to counselling careers with Artificial Intelligence: A systematic literature review}.
\newblock \bibinfo{journal}{\emph{Ricerche di Pedagogia e Didattica. Journal of Theories and Research in Education}} \bibinfo{volume}{18}, \bibinfo{number}{3} (\bibinfo{year}{2023}), \bibinfo{pages}{143--164}.
\newblock


\bibitem[Paas et~al\mbox{.}(2003)]%
        {paas_cognitive_2003}
\bibfield{author}{\bibinfo{person}{Fred Paas}, \bibinfo{person}{Juhani~E. Tuovinen}, \bibinfo{person}{Huib Tabbers}, {and} \bibinfo{person}{Pascal W.~M. Van~Gerven}.} \bibinfo{year}{2003}\natexlab{}.
\newblock \showarticletitle{Cognitive {Load} {Measurement} as a {Means} to {Advance} {Cognitive} {Load} {Theory}}.
\newblock \bibinfo{journal}{\emph{Educational Psychologist}} \bibinfo{volume}{38}, \bibinfo{number}{1} (\bibinfo{date}{Jan.} \bibinfo{year}{2003}), \bibinfo{pages}{63--71}.
\newblock
\showISSN{0046-1520}
\href{https://doi.org/10.1207/S15326985EP3801_8}{doi:\nolinkurl{10.1207/S15326985EP3801_8}}
\newblock
\shownote{Publisher: Routledge \_eprint: https://doi.org/10.1207/S15326985EP3801\_8}.


\bibitem[Pancholi et~al\mbox{.}(2009)]%
        {pancholi2009}
\bibfield{author}{\bibinfo{person}{Bhavna Pancholi}, \bibinfo{person}{Mark Dunne}, {and} \bibinfo{person}{Richard Armstrong}.} \bibinfo{year}{2009}\natexlab{}.
\newblock \showarticletitle{Sample size estimation and statistical power analyses}.
\newblock   \bibinfo{volume}{16} (\bibinfo{date}{11} \bibinfo{year}{2009}).
\newblock


\bibitem[Pandian and Suleri(2020)]%
        {pandian_nasa-tlx_2020}
\bibfield{author}{\bibinfo{person}{Vinoth Pandian~Sermuga Pandian} {and} \bibinfo{person}{Sarah Suleri}.} \bibinfo{year}{2020}\natexlab{}.
\newblock \bibinfo{title}{{NASA}-{TLX} {Web} {App}: {An} {Online} {Tool} to {Analyse} {Subjective} {Workload}}.
\newblock
\urldef\tempurl%
\url{http://arxiv.org/abs/2001.09963}
\showURL{%
\tempurl}
\newblock
\shownote{arXiv:2001.09963 [cs]}.


\bibitem[Pilli(2024)]%
        {pilli_exploring_2024}
\bibfield{author}{\bibinfo{person}{Stephen Pilli}.} \bibinfo{year}{2024}\natexlab{}.
\newblock \bibinfo{title}{Exploring {Conversational} {Agents} as an {Effective} {Tool} for {Measuring} {Cognitive} {Biases} in {Decision}-{Making}}.
\newblock
\href{https://doi.org/10.48550/arXiv.2401.06686}{doi:\nolinkurl{10.48550/arXiv.2401.06686}}
\newblock
\shownote{arXiv:2401.06686 [cs]}.


\bibitem[Rastogi et~al\mbox{.}(2020)]%
        {rastogi_towards_2020}
\bibfield{author}{\bibinfo{person}{Abhinav Rastogi}, \bibinfo{person}{Xiaoxue Zang}, \bibinfo{person}{Srinivas Sunkara}, \bibinfo{person}{Raghav Gupta}, {and} \bibinfo{person}{Pranav Khaitan}.} \bibinfo{year}{2020}\natexlab{}.
\newblock \showarticletitle{Towards {Scalable} {Multi}-{Domain} {Conversational} {Agents}: {The} {Schema}-{Guided} {Dialogue} {Dataset}}.
\newblock \bibinfo{journal}{\emph{Proceedings of the AAAI Conference on Artificial Intelligence}} \bibinfo{volume}{34}, \bibinfo{number}{05} (\bibinfo{date}{April} \bibinfo{year}{2020}), \bibinfo{pages}{8689--8696}.
\newblock
\showISSN{2374-3468}
\href{https://doi.org/10.1609/aaai.v34i05.6394}{doi:\nolinkurl{10.1609/aaai.v34i05.6394}}
\newblock
\shownote{Number: 05}.


\bibitem[Samuelson and Zeckhauser(1988)]%
        {samuelson_status_1988}
\bibfield{author}{\bibinfo{person}{William Samuelson} {and} \bibinfo{person}{Richard Zeckhauser}.} \bibinfo{year}{1988}\natexlab{}.
\newblock \showarticletitle{Status quo bias in decision making}.
\newblock \bibinfo{journal}{\emph{J Risk Uncertainty}} \bibinfo{volume}{1}, \bibinfo{number}{1} (\bibinfo{date}{March} \bibinfo{year}{1988}), \bibinfo{pages}{7--59}.
\newblock
\showISSN{1573-0476}
\href{https://doi.org/10.1007/BF00055564}{doi:\nolinkurl{10.1007/BF00055564}}


\bibitem[Schmidhuber et~al\mbox{.}(2021)]%
        {schmidhuber_cognitive_2021}
\bibfield{author}{\bibinfo{person}{Johanna Schmidhuber}, \bibinfo{person}{Stephan Schlögl}, {and} \bibinfo{person}{Christian Ploder}.} \bibinfo{year}{2021}\natexlab{}.
\newblock \showarticletitle{Cognitive {Load} and {Productivity} {Implications} in {Human}-{Chatbot} {Interaction}}. In \bibinfo{booktitle}{\emph{2021 {IEEE} 2nd {International} {Conference} on {Human}-{Machine} {Systems} ({ICHMS})}}. \bibinfo{pages}{1--6}.
\newblock
\href{https://doi.org/10.1109/ICHMS53169.2021.9582445}{doi:\nolinkurl{10.1109/ICHMS53169.2021.9582445}}


\bibitem[Sweller(1988)]%
        {sweller1988cognitive}
\bibfield{author}{\bibinfo{person}{John Sweller}.} \bibinfo{year}{1988}\natexlab{}.
\newblock \showarticletitle{Cognitive load during problem solving: Effects on learning}.
\newblock \bibinfo{journal}{\emph{Cognitive science}} \bibinfo{volume}{12}, \bibinfo{number}{2} (\bibinfo{year}{1988}), \bibinfo{pages}{257--285}.
\newblock


\bibitem[Sweller(2011)]%
        {sweller_chapter_2011}
\bibfield{author}{\bibinfo{person}{John Sweller}.} \bibinfo{year}{2011}\natexlab{}.
\newblock \showarticletitle{{CHAPTER} {TWO} - {Cognitive} {Load} {Theory}}.
\newblock In \bibinfo{booktitle}{\emph{Psychology of {Learning} and {Motivation}}}, \bibfield{editor}{\bibinfo{person}{Jose~P. Mestre} {and} \bibinfo{person}{Brian~H. Ross}} (Eds.). Vol.~\bibinfo{volume}{55}. \bibinfo{publisher}{Academic Press}, \bibinfo{pages}{37--76}.
\newblock
\href{https://doi.org/10.1016/B978-0-12-387691-1.00002-8}{doi:\nolinkurl{10.1016/B978-0-12-387691-1.00002-8}}


\bibitem[Thaler et~al\mbox{.}(2010)]%
        {thaler_choice_2010}
\bibfield{author}{\bibinfo{person}{Richard~H. Thaler}, \bibinfo{person}{Cass~R. Sunstein}, {and} \bibinfo{person}{John~P. Balz}.} \bibinfo{year}{2010}\natexlab{}.
\newblock \bibinfo{title}{Choice {Architecture}}.
\newblock
\href{https://doi.org/10.2139/ssrn.1583509}{doi:\nolinkurl{10.2139/ssrn.1583509}}


\bibitem[Tversky and Kahneman(1974)]%
        {tversky_judgment_1974}
\bibfield{author}{\bibinfo{person}{Amos Tversky} {and} \bibinfo{person}{Daniel Kahneman}.} \bibinfo{year}{1974}\natexlab{}.
\newblock \showarticletitle{Judgment under {Uncertainty}: {Heuristics} and {Biases}}.
\newblock   \bibinfo{volume}{185} (\bibinfo{year}{1974}).
\newblock


\bibitem[Van~Eyghen(2022)]%
        {VanEyghen_2022}
\bibfield{author}{\bibinfo{person}{Hans Van~Eyghen}.} \bibinfo{year}{2022}\natexlab{}.
\newblock \showarticletitle{Cognitive Bias: Phylogenesis or Ontogenesis?}
\newblock \bibinfo{journal}{\emph{Frontiers in Psychology}}  \bibinfo{volume}{13} (\bibinfo{year}{2022}).
\newblock
\showISSN{1664-1078}
\href{https://doi.org/10.3389/fpsyg.2022.892829}{doi:\nolinkurl{10.3389/fpsyg.2022.892829}}


\bibitem[Wu et~al\mbox{.}(2020)]%
        {wu_tod-bert_2020}
\bibfield{author}{\bibinfo{person}{Chien-Sheng Wu}, \bibinfo{person}{Steven~C.H. Hoi}, \bibinfo{person}{Richard Socher}, {and} \bibinfo{person}{Caiming Xiong}.} \bibinfo{year}{2020}\natexlab{}.
\newblock \showarticletitle{{TOD}-{BERT}: {Pre}-trained {Natural} {Language} {Understanding} for {Task}-{Oriented} {Dialogue}}. In \bibinfo{booktitle}{\emph{Proceedings of the 2020 {Conference} on {Empirical} {Methods} in {Natural} {Language} {Processing} ({EMNLP})}}. \bibinfo{publisher}{Association for Computational Linguistics}, \bibinfo{address}{Online}, \bibinfo{pages}{917--929}.
\newblock
\href{https://doi.org/10.18653/v1/2020.emnlp-main.66}{doi:\nolinkurl{10.18653/v1/2020.emnlp-main.66}}


\bibitem[Xiao et~al\mbox{.}(2021)]%
        {xiao_revisiting_2021}
\bibfield{author}{\bibinfo{person}{Qinyu Xiao}, \bibinfo{person}{Emma Lam}, \bibinfo{person}{Muhrajan Piara}, {and} \bibinfo{person}{Gilad Feldman}.} \bibinfo{year}{2021}\natexlab{}.
\newblock \showarticletitle{Revisiting status quo bias: {Replication} of {Samuelson} and {Zeckhauser} (1988)}.
\newblock \bibinfo{journal}{\emph{Meta-Psychology}}  \bibinfo{volume}{5} (\bibinfo{date}{Feb.} \bibinfo{year}{2021}).
\newblock
\href{https://doi.org/10.15626/MP.2020.2470}{doi:\nolinkurl{10.15626/MP.2020.2470}}


\bibitem[Yamamoto(2024)]%
        {yamamoto_suggestive_2024}
\bibfield{author}{\bibinfo{person}{Yusuke Yamamoto}.} \bibinfo{year}{2024}\natexlab{}.
\newblock \showarticletitle{Suggestive answers strategy in human-chatbot interaction: a route to engaged critical decision making}.
\newblock \bibinfo{journal}{\emph{Frontiers in Psychology}}  \bibinfo{volume}{15} (\bibinfo{date}{March} \bibinfo{year}{2024}), \bibinfo{pages}{1382234}.
\newblock
\showISSN{1664-1078}
\href{https://doi.org/10.3389/fpsyg.2024.1382234}{doi:\nolinkurl{10.3389/fpsyg.2024.1382234}}


\bibitem[Zhang et~al\mbox{.}(2023)]%
        {zhang_sgp-tod_2023}
\bibfield{author}{\bibinfo{person}{Xiaoying Zhang}, \bibinfo{person}{Baolin Peng}, \bibinfo{person}{Kun Li}, \bibinfo{person}{Jingyan Zhou}, {and} \bibinfo{person}{Helen Meng}.} \bibinfo{year}{2023}\natexlab{}.
\newblock \showarticletitle{{SGP}-{TOD}: {Building} {Task} {Bots} {Effortlessly} via {Schema}-{Guided} {LLM} {Prompting}}. In \bibinfo{booktitle}{\emph{Findings of the {Association} for {Computational} {Linguistics}: {EMNLP} 2023}}. \bibinfo{publisher}{Association for Computational Linguistics}, \bibinfo{address}{Singapore}, \bibinfo{pages}{13348--13369}.
\newblock
\href{https://doi.org/10.18653/v1/2023.findings-emnlp.891}{doi:\nolinkurl{10.18653/v1/2023.findings-emnlp.891}}


\bibitem[Zhao et~al\mbox{.}(2023)]%
        {zhao_anytod_2023}
\bibfield{author}{\bibinfo{person}{Jeffrey Zhao}, \bibinfo{person}{Yuan Cao}, \bibinfo{person}{Raghav Gupta}, \bibinfo{person}{Harrison Lee}, \bibinfo{person}{Abhinav Rastogi}, \bibinfo{person}{Mingqiu Wang}, \bibinfo{person}{Hagen Soltau}, \bibinfo{person}{Izhak Shafran}, {and} \bibinfo{person}{Yonghui Wu}.} \bibinfo{year}{2023}\natexlab{}.
\newblock \bibinfo{title}{{AnyTOD}: {A} {Programmable} {Task}-{Oriented} {Dialog} {System}}.
\newblock
\urldef\tempurl%
\url{http://arxiv.org/abs/2212.09939}
\showURL{%
\tempurl}
\newblock
\shownote{arXiv:2212.09939 [cs]}.


\end{thebibliography}

\end{document}